\documentclass[twocolumn,amsmath,amssymb,superscriptaddress,prl,a4paper]{revtex4}
\usepackage{graphicx}
\usepackage[paperwidth=210mm,paperheight=297mm,centering,hmargin=2.0cm,vmargin=2.6cm]{geometry}

\renewcommand{\vec}[1]{\boldsymbol{#1}}
\newcommand{\beq}{\begin{equation}}
\newcommand{\eeq}{\end{equation}}
\newcommand{\beqa}{\begin{eqnarray}}
\newcommand{\eeqa}{\end{eqnarray}}
\newcommand{\e}{\mathrm{e}}
\newcommand{\w}{\omega}

\newcommand{\ket}[1]{\left| #1 \right\rangle}
\newcommand{\bra}[1]{\left\langle #1 \right|}

\newcommand{\av}[1]{\langle #1\rangle}
\newcommand{\braket}[2]{\langle #1 | #2\rangle}

\newcommand{\ketbra}[2]{\left|#1\right\rangle\hskip-1mm\left\langle #2\right|}

\begin{document}

\title{Error distributions on large entangled states with non-Markovian dynamics}
\author{Dara P. S. McCutcheon}
\affiliation{Blackett Laboratory, Imperial College London, London SW7 2AZ, UK}
\affiliation{{Departamento de F\'isica, FCEyN, UBA and IFIBA, Conicet,~Pabell\'on 1, Ciudad Universitaria, 1428 Buenos Aires, Argentina}}
\affiliation{Department of Photonics Engineering, DTU Fotonik, \O rested Plads, 2800 Kgs. Lyngby, Denmark}
\author{Netanel H. Lindner}
\affiliation{Department of Physics,Technion - Israel Institute of Technology, Haifa 32000, Israel}
\author{Terry Rudolph}
\affiliation{Blackett Laboratory, Imperial College London, London SW7 2AZ, UK}

\date{\today}

\begin{abstract}

We investigate the distribution of errors on a computationally useful entangled state generated via the repeated emission
from an emitter undergoing strongly non-Markovian evolution. For emitter-environment coupling of pure-dephasing form,
we show that the probability that a particular patten of errors occurs has a bound of
Markovian form, and thus accuracy threshold
theorems based on Markovian models should be just as effective. 
Beyond the pure-dephasing assumption, though
complicated error structures can arise, they can still be qualitatively bounded by a Markovian error model.

\end{abstract}

\maketitle

The theoretical and technical challenges faced in the construction of a quantum computer have
rightly brought into light the question of the scalability of such a device~\cite{unruh95,alicki02,alicki06,alickiarxivs,kalaiarxivs}.
There is, however, cause for optimism, since accuracy threshold theorems imply
that quantum computation should be achievable to arbitrary
precision~\cite{aharonov98,knill98,preskill98,aliferis05,terhal05,aharonov06,preskill12}. The existence of such thresholds relies on
quantum error correction codes~\cite{shor95,steane96,laflamme96}, and that the noise afflicting the computation device satisfies
certain conditions on its strength, and level of spatial and temporal correlations. In particular,
the first theoretical achievements assumed Markovian and independent noise afflicting the components
of the quantum computer~\cite{aharonov98,knill98,preskill98}, the intuition being that, typically,
the components of the device reside in different locations,
and that the (local) environments causing the errors are large enough that they have
an effectively negligible memory time~\cite{b+p}.

While these assumptions will be valid in some quantum systems, whether they are valid for those systems 
able to perform quantum computations remains to be seen.
Recently, prompted by a debate between Kalai and Harrow,
considerable discussion has taken place in the community about some of the core assumptions of the error models underpinning threshold
theorems for fault tolerant quantum computing~\cite{kalaiarxivs,preskill12,blog,harrow12}. Broadly speaking,
questions have been raised about the spatial and temporal structure of errors incurred when one
creates large entangled states without the usual assumption of Markovian dynamics.

The purpose of this work is twofold: In the first instance, we analyse a
worst-case scenario, wherein a large photonic cluster state~\cite{raussendorf01,raussendorf03}
is created by a single emitter that is continuously coupled to an
environment in a highly non-Markovian manner. In this scenario, it is reasonable to believe that all the
errors on this photonic state arise from the emission process,
as once the photons are travelling in free space they are effectively
decoherence-free. Secondly, we analyse this procedure
in the context of a specific experimental proposal with realistic parameters, where the emitter is a charged quantum dot
interacting with a nuclear spin bath. 
As our main result, we show that one can obtain a bound on the non-Markovian error distribution
probabilities which has a Markovian form.  Crucially, this means that methods for combating
Markovian errors will work just as efficiently in this highly non-Markovian situation. 
When the emitter is subject to pure-dephasing noise - as can be the case for the 
proposal we consider - our bound is analytically derived. Outside this regime, the structure of 
errors becomes more complicated, though we 
show numerically how a Markovian error model can still correctly capture all features and provide a tight bound.

\begin{figure}
\begin{center}
\includegraphics[width=0.35\textwidth]{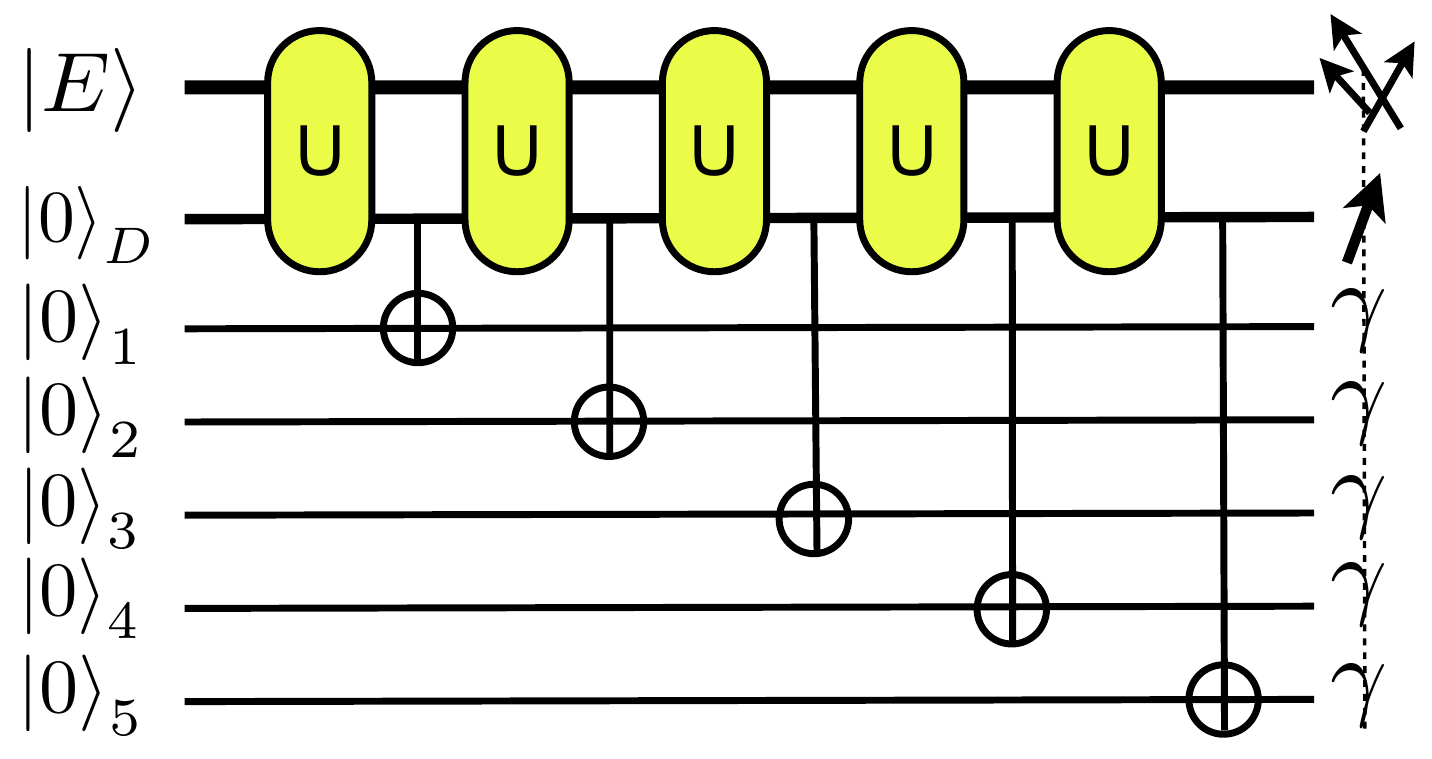}
\caption{Quantum circuit for the generation of a linear cluster state of five photons and a quantum dot. Non-Markovian decoherence of
the dot is modelled as a sequential coupling to an environment. In the ideal case the unitaries are replaced by $\e^{-i Y_D \pi/4}\otimes\openone_E$.}
\label{circuit}
\end{center}
\end{figure}

In order to give a context, we phrase our augments in terms of the linear
cluster proposal of Ref.~\cite{lindner09}, which consists of the
repeated absorption and reemission of a string of photons from a quantum dot (QD) residing in a magnetic field
perpendicular to the growth direction~\footnote{While linear cluster states are not
sufficient for universal computation, using linear optics alone they may be fused into higher dimension structures~\cite{browne05},
and it seems implausible that a complicated error structure could accumulate owing to the fusion process.}.
We note that there are practical proposals
(with experiments underway) to build such devices~\cite{lindner09,li11,sophia10,lin10}, though we emphasise that
our analysis equally applies to any cluster state produced in a similar manner.
In the ideal case (no coupling to an environment), a state of $n$ entangled photons and the QD, $\ket{\mathcal{C}_n}$,
is generated from an initially separable state via [see Fig.~({\ref{circuit}})]:
\beq
\ket{\mathcal{C}_n}=\left[\prod_{i=1}^n (C_i U_y)\right]\ket{0}_{D}\ket{0\dots0}
\label{CIdeal}
\eeq
where $\ket{0}_D\equiv\ket{\uparrow}$ ($\ket{1}_D\equiv\ket{\downarrow}$) is the state of the QD aligned (anti-aligned) along the z-axis,
$\ket{0\dots0}\equiv\bigotimes_{i=1}^n\ket{R}$
represents the initial state of the $n$-photons all having right circular polarisation, $C_i=\ketbra{0}{0}_D\otimes\openone_i+\ketbra{1}{1}_D\otimes X_i$ is a CNOT gate
on the QD and the ith photon representing an absorption and emission process, and $U_y=\e^{-i Y_D \pi/4}$ rotates the QD about the y-axis. Our basis is such that
$Z_D=\ketbra{0}{0}_D-\ketbra{1}{1}_D$ and $Z_i=\ketbra{0}{0}_i-\ketbra{1}{1}_i$ where $\ket{1}_i\equiv\ket{L}_i$.
Non-Markovian evolution of the QD is introduced by sequentially coupling it to an environment such that $U_y\to U$, 
with $U$ acting on the QD and its environment.

Before we do so, we first simply consider the effect of Pauli errors on the QD before, say, the emission of photon $l$,
i.e. we insert $X_D$, $Y_D$, or $Z_D$ to the left of $C_{l-1}$ in
Eq.~({\ref{CIdeal}}). We refer to this type of error (on the QD itself as apposed to the resulting photon state) 
as a \emph{fundamental} error. 
We find that $\ket{\mathcal{C}_n}$ becomes
$Z_{l}\ket{\mathcal{C}_n}$, $Z_{l}Z_{l-1}\ket{\mathcal{C}_n}$, and $Z_{l-1}\ket{\mathcal{C}_n}$, for $X_D$, $Y_D$, and $Z_D$ 
respectively~\footnote{To see this most clearly one must insert $Z_l$ operators to the right of the 
$C_l$ operators in Eq.~({\ref{CIdeal}}), which have no effect on the initial state.}.
Thus, we see that imperfections in the evolution of the QD (fundamental errors),
are mathematically equivalent to \emph{localised} errors on the resulting photon state~\cite{lindner09}.

We now investigate how these errors are distributed.
We assume that the absorption and emission processes of the photons occur on a timescale far shorter than the rotations of the QD,
and the CNOT gates are therefore treated as being instantaneous. It was shown in Ref.~\cite{lindner09} that relaxation of this assumption gives rise
to photons with wave-packets which correspond to a fixed probability of a fundamental Y error on the QD for each CNOT gate. 
To model the non-Markovian evolution of the QD 
we replace $U_y$ in Eq.~({\ref{CIdeal}}) with the 
general operator $U=\ketbra{0}{0}_D A_{00}+\ketbra{0}{1}_D A_{01}+\ketbra{1}{0}_D A_{10}+\ketbra{1}{1}_D A_{11}$,
where the operators $A_{00}$ etc. act on the environment $\ket{E}$. Eq.~({\ref{CIdeal}}) becomes
$\ket{\Psi_n}=\left[\prod_{i=1}^n (C_i U)\right]\ket{0}_{D}\ket{0\dots0}\ket{E}$, which inspection reveals can be written
\begin{align}
\ket{\Psi_n}
=\sum_{\vec{b}}\ket{b_n}_D\ket{\vec{b}}\mathcal{F}(\vec{b})\ket{E}
\label{Psin}
\end{align}
where $\vec{b}=(b_n b_{n-1}\dots b_2 b_1)$ with $b_i\in\{0,1\}$ a bit string,
$\mathcal{F}(\vec{b})=A_{b_n b_{n-1}}A_{b_{n-1} b_{n-2}}\dots A_{b_2 b_1}A_{b_1 0}$ is a product of environment operators,
and the sum runs over all $2^n$ possible bit-strings $\vec{b}$. Eq.~({\ref{Psin}}) is the complete state of the QD,
$n$ photons and environment. Now, we denote by
$P(\vec{\alpha})$, where $\vec{\alpha}=(\alpha_n\alpha_{n-1}\dots\alpha_2\alpha_1)$ with $\alpha_i \in \{0,1\}$, the probability
that the photonic state is measured having Pauli $Z$ errors on those photons for which $\alpha_i=1$,
i.e. the state $\ket{\Phi(\vec{\alpha})}=[\prod_{i=1}^n Z_{i}^{\alpha_i}]\ket{\mathcal{C}_n}$. 
We find $P(\vec{\alpha})=\bra{\Phi(\vec{\alpha})}{\rm{Tr}}_E(\ketbra{\Psi_n}{\Psi_n})\ket{\Phi(\vec{\alpha})}=
\mathrm{Tr}_E(O(\vec{\alpha})^{\dagger}O(\vec{\alpha})\ketbra{E}{E})$, where
the environment operator $O(\vec{\alpha})=\sqrt{2}\bra{+}_DW(\vec{\alpha})\ket{0}_D$ is a matrix element
of the QD--environment operator
\begin{align}
W(\vec{\alpha})=\prod_{i=1}^n (Z^{\alpha_i}\Delta)
=(Z^{\alpha_n}\Delta)\dots(Z^{\alpha_1}\Delta),
\label{WAlpha}
\end{align}
with $\ket{+}_D=(1/\sqrt{2})(\ket{0}_D+\ket{1}_D)$, $Z=Z_D\otimes \openone_E$, and
$\Delta=(1/\sqrt{2})(U-2\ketbra{0}{0}_D U \ketbra{1}{1}_D)$ is a non-unitary operator acting in the joint QD-environment Hilbert space.
For the probability of zero errors, for example, we have the scalar $P(\vec{0})=\mathrm{Tr}_E(O(\vec{0})^{\dagger}O(\vec{0})\ketbra{E}{E})$,
which depends on the environment operator $O(\vec{0})=\sqrt{2}\bra{+}_D W(\vec{0})\ket{0}_D$,
which, from Eq.~({\ref{WAlpha}}), in turn depends on the
QD-environment operator $W(\vec{0})=\Delta^n$.
For the probability of an error on, say, photon $l$, the relevant operator is
$W(0\dots010\dots0)=\Delta^{n-l}Z\Delta^l$, and so on. Thus, calculating the probability of a given error distribution amounts to
calculating products of $Z$ and the non-Hermitian matrix 
$\Delta$. Eq.~({\ref{WAlpha}}) provides us with a systematic way to 
determine error distribution probabilities in the non-Markovian case, making no assumptions
about the state of the environment, its memory timescale, or its interaction strength with, or potential correlations with,
the QD at any point in the evolution. 
Though we have phrased our analysis in terms of quantum dots and photons, Eq.~({\ref{WAlpha}}) is valid for any cluster state generated
as shown in Fig.~({\ref{circuit}}).

For emitter--environment coupling of pure-dephasing form, Eq.~({\ref{WAlpha}}) can be further simplified. 
We motivate this by noting that 
for electrons in QDs, the dominant
source of dephasing is due to coupling to nuclear spins via hyper-fine
interactions~\cite{luka09,luka09prl,coish10,barnes12}. Since we consider a field in the $y$-direction, the Hamiltonian 
takes the form 
$H=(\Omega/2)Y_D+(1/2)\sum_k \w_k I_k^{y}+(1/4)\sum_k A_k \vec{S}\cdot \vec{I}_k+H_{\mathrm{dip}}$, where 
$\vec{S}=(X_D,Y_D,Z_D)$, while $\vec{I}_k=(I_k^x,I_k^y,I_k^z)$ acts on environment spin $k$, and 
$H_{\mathrm{dip}}=\sum_{k\neq k'} b_{k k'}(I_k^+ I_{k'}^--(1/2)I_k^y I_{k'}^y)$
with $I_k^{\pm}=(1/2)(I_k^x\pm i I_k^z)$. 
Typically, the Zeeman energy of the QD spin is far larger than those of the nuclei, leading to a suppression of relaxation
processes. The quantity regulating this distinction is
$\delta = \mathcal{A}/(\Omega\sqrt{N}$), where $\mathcal{A}=\sum_k A_k$ and $N$ is the number of nuclei appreciably
interacting with the QD spin. For $\delta\ll 1$, it was shown that the full Hamiltonian above can be approximated by
the pure-dephasing Hamiltonian~\cite{luka09,luka09prl} 
$H_{\mathrm{PD}}=(\Omega_{\mathrm{eff}}/2)Y_D+\ketbra{+i}{+i}H_++\ketbra{-i}{-i}H_-$ where $Y_D\ket{\pm i}=\pm\ket{\pm i}$ and
$H_{\pm}=\mp(1/2)\av{B_N}+(1/2)\sum_k (\w_k'\pm{\textstyle{\frac{1}{2}}}A_k)I_k^y
\pm(1/4)\sum_{k\neq k'} \frac{A_k A_{k'}}{\Omega}I_k^+ I_{k'}^-+H_{\mathrm{dip}}$, 
with $\w_k'=\w_k-A_k^2/(4\Omega)$. The effective magnetic field is $\Omega_{\mathrm{eff}}=\Omega+\av{B_N}+(1/4)\sum_k A_k^2/\Omega$ 
where $\av{B_N}=\mathrm{Tr}(B_N \rho_E)$ with $B_N=(1/2)\sum_k A_k I_k^y$ the Overhauser field operator. 
For typical GaAs QDs the total coupling strength $\mathcal{A}\sim 1~$T, while the typical values of
N range from $10^4$ to $10^6$~\cite{luka09,luka09prl,coish10}. Thus, field strengths of $\Omega\sim 100$~mT and above
should be well described by the pure-dephasing Hamiltonian.

Using $U=\exp[-i \pi/(2\Omega_{\mathrm{eff}})H_{\mathrm{PD}}]$, from Eq.~({\ref{WAlpha}}) we find for
a general error distribution $\vec{\alpha}$ we have
\beq
O(\vec{\alpha})=
\prod_{i=1}^n\frac{\sqrt{(-1)^{f_i(\vec{\alpha})}}}{2}\Big[U_-+(-1)^{f_i(\vec{\alpha})} U_+\Big]
\label{OPureDephasing}
\eeq
with $f_i(\vec{\alpha})=\sum_{j=i}^n\alpha_j$ and $U_{\pm}=\exp[-i (\pi/2\Omega_{\mathrm{eff}})H_{\pm}]$. We see that 
$O(\vec{\alpha})$ consists of a product of $n$ operators, each of which being either $M_{\pm}=(\sqrt{\pm1}/2)(U_-\pm U_+)$
depending on the error distribution $\vec{\alpha}$. Using Eq.~({\ref{OPureDephasing}}), and
the sub-multiplacative property of the operator norm
defined as $|| A ||\equiv \mathrm{max}_{\ket{\phi}}\bra{\phi} A \ket{\phi}/\braket{\phi}{\phi}$,
we find the non-Markovian error probabilities satisfy
\beq
P(\vec{\alpha})\leq|| M_-^{\dagger}M_-||^{h(\vec{\alpha})}|| M_+^{\dagger} M_+ ||^{n-h(\vec{\alpha})}
\label{PAlphaPD}
\eeq
where $h(\vec{\alpha})=\frac{1}{2}(n-\sum_{i=1}^n (-1)^{f_i(\vec{\alpha})})$ is the number of occurrences of $M_-$ in
Eq.~({\ref{OPureDephasing}}). We see that $||M_-^{\dagger}M_-||$ plays the role of an error probability, with unitarity of
$U_{\pm}$ ensuring that $|| M_{\pm}^{\dagger}M_{\pm}||\leq1$.
Note that $h(\vec{\alpha})$ does not count the number of errors on the
photonic state: it counts the number of \emph{adjacent pairs} necessary to create it, or equivalently, the number of
fundamental QD errors. The form of $H_{\mathrm{PD}}$ means that the environment can only induce
fundamental Y errors on the QD, which make 
adjacent pairs of errors on the resulting photonic cluster state. A single isolated error, say
$\vec{\alpha}=(010)$, has $h(010)=2$, 
since pairs of adjacent errors at positions $1$ and $2$ are required to realise it.

Eq.~({\ref{PAlphaPD}}) shows that even in the non-Markovian case,
we can put a rigorous bound on the probability of a given error distribution, which
behaves as a power law in the number of fundamental errors in the distribution. More importantly, we see
that the non-Markovian nature of the environment cannot introduce long range spatial correlations in the errors; the probability of
$h$ fundamental errors is bounded by $p_-^h$ with $\smash{p_-=||M_-^{\dagger}M_-||\leq1}$.
These results are valid for any cluster state generated in the way shown in Fig.~({\ref{circuit}}) when 
the emitter-environment coupling takes on a pure-dephasing form.

For the QD example we consider, we can go further by noticing 
that the total spin projection in the $y$-direction is conserved, $[H_{\pm},\sum_k I_k^y]=0$. The operators 
$U_{\pm}=\exp[-i (\pi/2\Omega_{\mathrm{eff}}) H_{\pm}]$ 
from which the probabilities are calculated are therefore block-diagonal, and the result is that the probabilities become
a sum over contributions from spaces with fixed spin projections. 
By defining projection operators
$\mathcal{P}_m$ which satisfy $\sum_m \mathcal{P}_m=\openone_E$
and project onto the eigenspace with eigenvalue of $\sum_k I_k^y$ equal to $m$,
the probabilities can be written $P(\vec{\alpha})=\sum_m P_m(\vec{\alpha})$ where
$P_m(\vec{\alpha})=\mathrm{Tr}_E(O_m^{\dagger}(\vec{\alpha})O_m(\vec{\alpha}) \rho_m)$, and
$\rho_m$ is the environment state in the $m$ subspace, while $O(\vec{\alpha})=\mathcal{P}_m O(\vec{\alpha})\mathcal{P}_m$. 
In this way, we can make use of properties we know of the environment state. 
For example, for an initial environment state having weight in a single $m$ sector only,
we can write $P(\vec{\alpha})=P_{m}(\vec{\alpha})$ and bound by
\beq
P_{m}(\vec{\alpha})\leq|| M_-^{(m)\dagger}M_-^{(m)}||^{h(\vec{\alpha})}|| M_+^{(m)\dagger} M_+^{(m)}||^{n-h(\vec{\alpha})}
\label{Boundm}
\eeq
where $M_{\pm}^{(m)}=\mathcal{P}_m M_{\pm}\mathcal{P}_m$. This bound is tighter than that given in
Eq.~({\ref{PAlphaPD}}) since the operators involved necessarily act non-trivially in a smaller space.

\begin{figure}
\begin{center}
\includegraphics[width=0.48\textwidth]{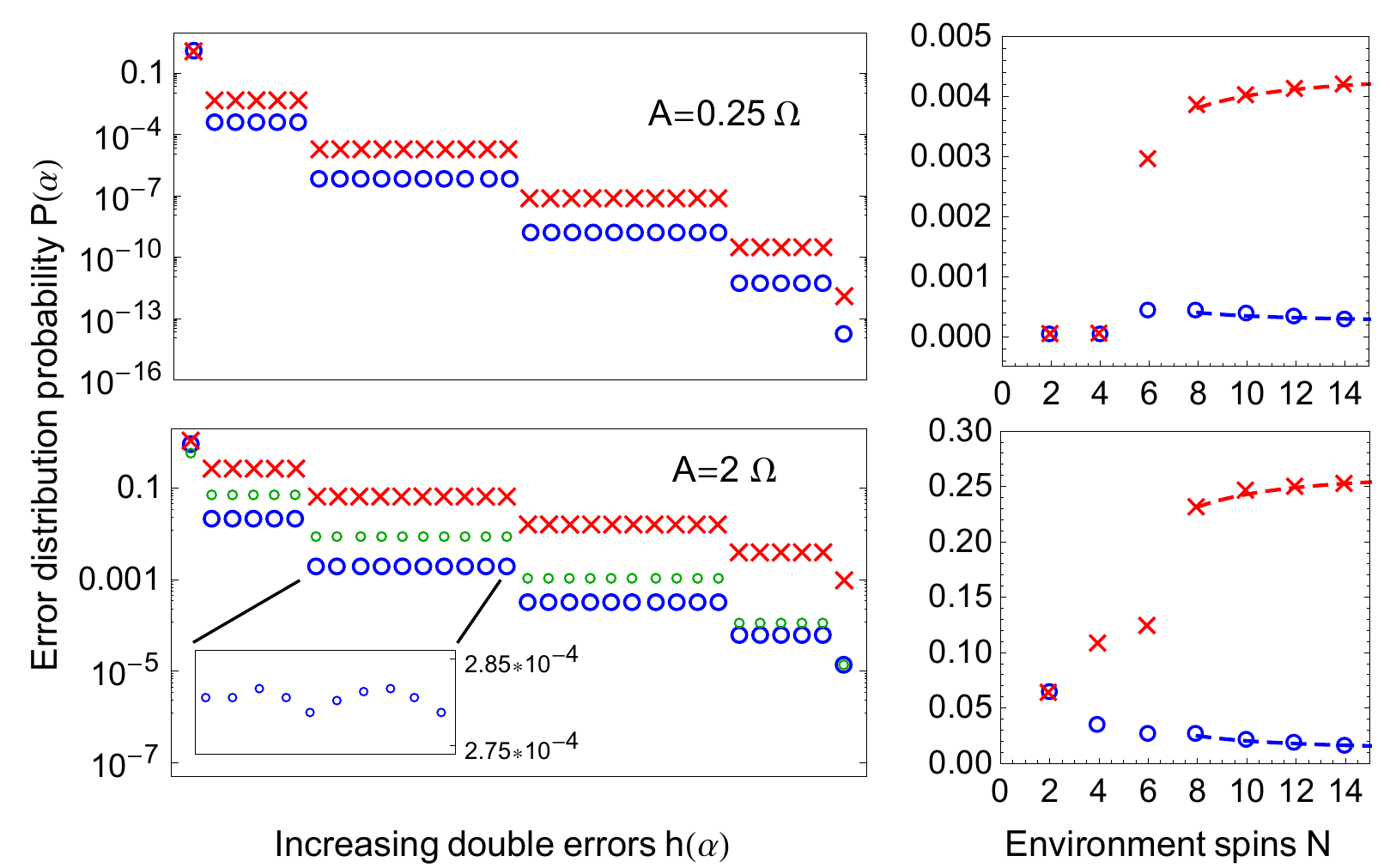}
\caption{Left panels: all $2^5$ non-Markovian error distribution probabilities for a 5-photon state, calculated
exactly (blue circles), using the bound given in Eq.~({\ref{Boundm}}) (red crosses), and a best-fit to a Markovian model of the form
$p^h(1-p)^{n-h}$ (green circles). The error distributions are ordered along the
$x$-axis such that those corresponding to the least number of fundamental errors are to the left. The inset in the lower plot shows a zoom in
of the $h(\vec{\alpha})=2$ band. Right panels: scaling of
a typical error distribution probability with increasing environment size.}
\label{ErrorPlots}
\end{center}
\end{figure}

In Fig.~({\ref{ErrorPlots}}) we plot the exact non-Markovian error probabilities (blue circles) and the bound
calculated using Eq.~({\ref{Boundm}}) (red crosses), using the pure-dephasing Hamiltonian. The left panels show all $2^5$ error 
probabilities for a $5$-photon state, ordered by increasing fundamental errors, $h(\vec{\alpha})$, for an environment
of $N=10$ spins initially in an equal mixture in the $m=0$ subspace~\footnote{The coupling coefficients follow a Gaussian distribution,
$A_k \propto \exp[-( 2k/N)^2]$ such that $\sum_{k=1}^N A_k = \mathcal{A}$ independent of $N$.
Similarly $b_{kk'}\propto \exp[-( 2k/N)^2-( 2k'/N)^2]$ such that $\sum_{k'=1}^N b_{kk'}/A_k = 2500$, while
$\sum_k \w_k'/\Omega_{\mathrm{eff}} =10^{-3}$.}.
The probabilities fall into distinct bands determined by 
their value of $h(\vec{\alpha})$, and the bounds correctly 
capture the behaviour of the exact values. For small $\mathcal{A}/\Omega$ our bound is relatively tight, 
while for $\mathcal{A}/\Omega =2$ where our derived bound gives fairly high values, the exact probabilities
are still well behaved and remain low. In fact, they can be bounded using a simple best-fit procedure by a
Markovian model of the form $p^h(1-p)^{n-h}$, with $p$ significantly less than $p_-$, as show in green on the lower left plot. 
It is clear from Fig.~({\ref{ErrorPlots}}) 
that the non-Markovian errors do not show harmful long-range correlations,
as the bound suggests. Thus, strategies to combat errors assuming Markovian evolution will, in this regime, remain effective 
in the non-Markovian case.

Note that we have chosen here an initial environment state for which the {\emph{average}} Overhauser field is zero $\av{B_N}=0$,
and for which fluctuations are small, $\Delta B_N = \sqrt{\langle (B_N)^2 \rangle - \langle B_N \rangle ^2}\ll \Omega$. 
As a result, dephasing due to ensemble average over the Zeeman field is eliminated, and the error probabilities 
remain low and \textit{approximately} equal.
Importantly, as long as the initial state obeys $\Delta B_N\ll\Omega$, the features seen in Fig.~(2) are remarkably robust; since we include
$\av{B_N}$ in $\Omega_{\mathrm{eff}}$, they are also present in cases for which
$\av{B_N}\neq 0$, including initially pure environment states.

In the right panels of Fig.~({\ref{ErrorPlots}}) we show the scaling with
increasing environment size
of the exact probability and the bound in Eq.~({\ref{Boundm}}), for a typical error distribution 
$\vec{\alpha}=(01100)$ for which $h(\vec{\alpha})=1$. 
With pure-dephasing, the exact probabilities ought to scale as $\sim 1- (1 + |a|\, N^{-2})^{-1/2}$ where $a$ is fixed
for fixed $\mathcal{A}$ and $\Omega$~\cite{luka09,luka09prl}, and we find that the bound obeys a similar scaling 
$\sim c-(1 - |a|\, N^{-2})^{-1/2}$.
The dashed lines show fits of this form, showing that the probabilities and bound scale
as expected with $N$. Thus, the bound we derive tends to a constant value with increasing environment size, and for small 
$\mathcal{A}/\Omega$, can directly replace the error rate in threshold theorems assuming \emph{Markovian} error models.
In fact, even when $\mathcal{A}/\Omega$ takes on higher values, our numerics strongly suggest that one can  \textit{tightly} bound the error distribution with a Markovian model.

\begin{figure}
\begin{center}
\includegraphics[width=0.48\textwidth]{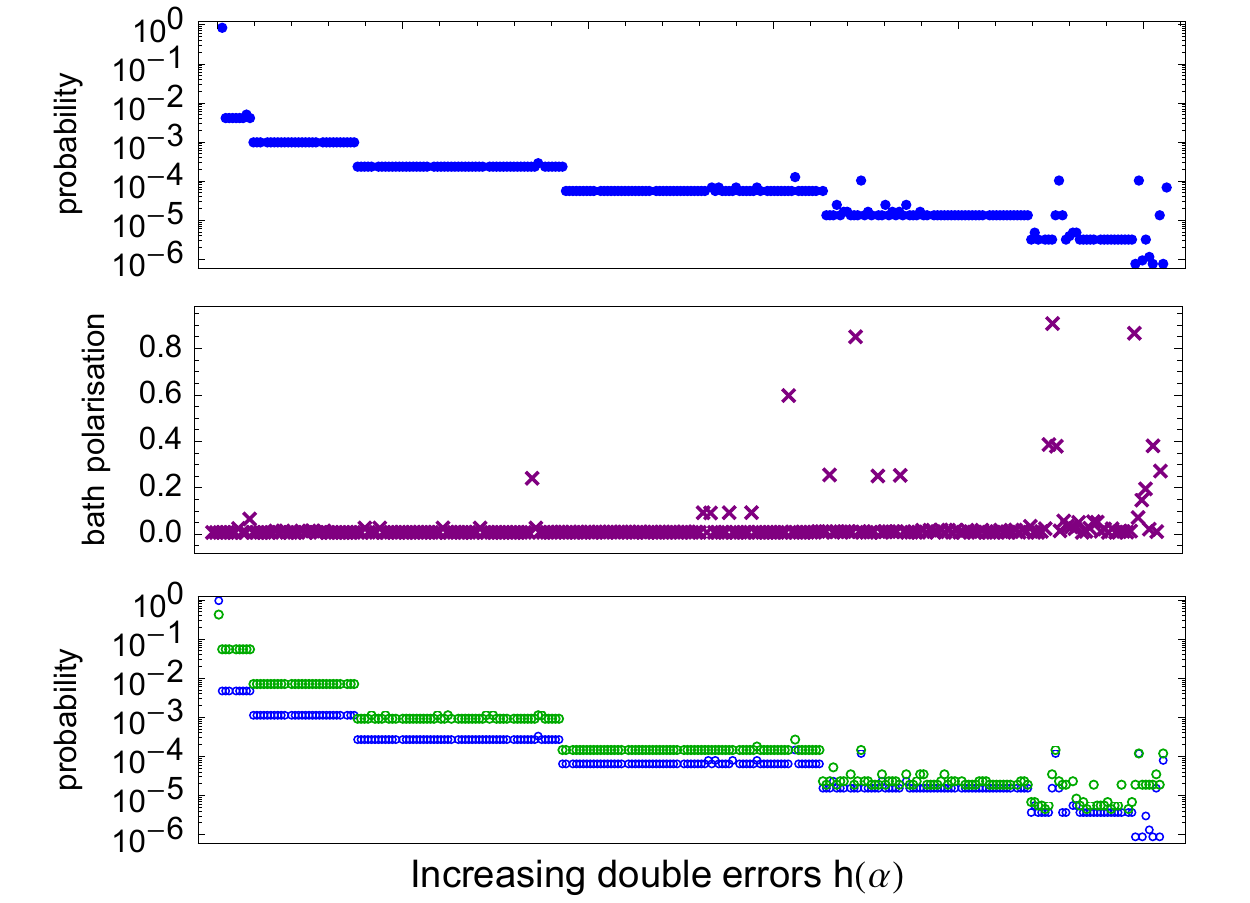}
\caption{Top: all $2^{8}$ exact non-Markovian error probabilities for a 8 photon state in the non-pure-dephasing case.
Middle: corresponding polarisation $\av{\sum_k I_k^y}$ of the spin environment: error probabilities falling outside their bands
correspond to angular momentum exchanges. Bottom: the qualitative features of the probability distribution can be captured by
a simple Markovian model shown in green.}
\label{ErrorPlotsPh10}
\end{center}
\end{figure}

For typical QDs the pure-dephasing form is valid for magnetic field strengths
of $\sim100$~mT and above. However, for optimal performance of the specific cluster state
proposal we consider~\cite{lindner09} smaller magnetic fields would be preferred (though not essential).
To investigate this regime, in the top panel of Fig.~({\ref{ErrorPlotsPh10}}) we
show the non-Markovian probabilities calculated using the full hyper-fine Hamiltonian,
for an 8 photon state with $\mathcal{A}/\Omega =4$ and $N=6$ such that $\delta\approx 1.6$ (so that $\Omega \sim 1~\mathrm{mT}$ for
realistic QD sizes). We see that the band structure becomes convolved with probabilities that lie above their bands. These
distributions all have the form $\vec{\alpha}_1=(\dots 1\dots)$; the
Hamiltonian we now use can induce fundamental X and Z errors on the QD,
which correspond to \emph{single} errors on the photon state. This can be further understood in the middle panel,
where we show the corresponding polarisation of the environment $\av{\sum_k I_k^y}$ for each distribution;
when a distribution of the form $\vec{\alpha}_1$ is realised, angular momentum is exchanged with the environment.

Though these exact non-Markovian probabilities appear to
have a more complicated structure, they can still be qualitatively described by a Markovian model.
With any error distribution $\vec{\alpha}$ we can associate a finite number of QD trajectories which will result in it. An error distribution
$\vec{\alpha}_1$, for example, can be made from a combination of fundamental Y errors, or a
single X or Z error. We can therefore define a simple Markovian model, wherein we assign fixed
probabilities for fundamental $X$, $Y$, and $Z$ errors, from these calculate the probability of a given trajectory, and
sum over all trajectories corresponding to a given error distribution. Probabilities calculated in this way are
shown in green in the lower panel of Fig.~({\ref{ErrorPlotsPh10}}). Importantly, we see that these Markovian error probabilities
qualitatively capture all the exact non-Markovian probabilities.

We have investigated the distribution of errors on a large entangled state generated by the repeated emission
from a single emitter with non-Markovian evolution. For pure-dephasing dynamics, we found that the error probabilities have
a bound of Markovian form, such that error correction schemes remain just as effective in this non-Markovian regime. We have 
also shown that the errors can be bounded by a Markovian model even beyond 
pure-dephasing dynamics, suggesting the board applicability of our findings.

{\it Acknowledgments} - 
The authors wish to thank Sophia Economou and John Preskill for numerous useful discussions. 
D.P.S.M. acknowledges CONICET and SIQUTE (Contract No. EXL02) of the European Metrology 
Research Programme (EMRP) for support. 
The EMRP is jointly funded by the EMRP participating countries with EURAMET and the European Union. 
N.H.L. thanks the Israel Excellence Centre ``Circle of Light''. 
T.R. is supported by the Vienna Science and Technology Fund (WWTF, Grant No. ICT 12-041), and 
the Army Research Office (ARO) Grant No. W911NF-14-1-0133. 
D.P.S.M. and T.R. also acknowledge support from the EPSRC and
CHIST-ERA project SSQN.


\end{document}